\documentclass{aa}
\usepackage{psfig}

\begin{document}

\title{Atmospheric dynamics and the mass loss process in red supergiant stars}

\author{
E. Josselin     \inst{1}
\and B. Plez   \inst{1}
}

\offprints{E. Josselin}

\institute{GRAAL, Universit\'e Montpellier II - ISTEEM, CNRS, Place Eugene Bataillon, 
F-34095 Montpellier Cedex, France \\
\email{josselin,plez@graal.univ-montp2.fr}
}

\titlerunning{Atmospheric dynamics and the mass-loss process in red supergiant stars}

\date{Received / accepted }

\abstract
{Red supergiant stars represent a key phase in the evolution 
of massive stars. 
Recent radiative hydrodynamic simulations suggest that their 
atmospheres may be the location of large-scale convective motions.}
{As supergiant convection is expected to generate 
supersonic motions and shocks, we seek constraints on these 
atmospheric motions and their possible relation with mass-loss rates. }
{ We present high-resolution, visible spectroscopy of a sample of 
red supergiants (spectral type M I) and analyse them with a tomographic 
technique. }
{We observe steep velocity gradients, characterising both upward and 
downward supersonic motions, which are time variable on time scales 
of a few hundred days. }
{These convective motions will generate turbulent pressure, which will strongly 
decrease the effective gravity. We suggest that this decrease, combined with radiative 
pressure on molecular lines, initiate the mass loss in red supergiant stars. }
\keywords{Supergiants -- Stars: atmospheres -- Stars: mass-loss}

\maketitle 

\section{Introduction}
Red supergiants (RSG) are evolved massive 
(10 $\la$ M$_{\rm init}$ $\la$ 40 M$_\odot$) stars, in the core helium-burning 
phase, preceding Wolf-Rayet stars and/or type II supernovae. 
Thanks to their high-peak infrared luminosity, red supergiants are potential 
powerful tracers of galactic  structure, efficiently probing regions of high 
interstellar extinction. They may also become extragalactic distance indicators, 
if their fundamental parameters are properly calibrated. 

This task is hampered by the difficulty of carrying out a proper analysis 
of RSG optical spectra. Because of their low effective temperature, 
these spectra are dominated by strong molecular bands (esp. TiO), 
which lead to an ill-defined continuum. Furthermore, spectral lines 
exhibit a strong broadening, which can be parameterized  by an ad-hoc 
high macroturbulent velocity of the order of 15 km~s$^{-1}$, indicating 
probable supersonic motions (Josselin \& Plez 2004). 
 In addition,  non-LTE line formation is suspected for electronic molecular 
transitions (Hinkle \& Lambert 1975) and may also affect atomic lines, 
regarding the low surface gravity  of RSG. 
RSG lower atmospheres are convectively unstable. Already 30 years ago 
Schwarzschild (1975) claimed that large-scale convective motions were to be 
expected, with only a few granules covering the surface. 

Josselin et al. (2000, hereafter paper I) analysed infrared 
photometric and millimeter spectroscopic observations of a sample 
of RSG. They showed 
in particular that the dust mass-loss rate is not correlated 
with luminosity. Furthermore, the molecular gas-to-dust ratio 
shows a very large scatter and is generally higher than what is 
observed towards asymptotic giant branch (hereafter AGB) stars. 
Finally, Josselin al. found strong indications of circumstellar 
inhomogeneities for at least one object, VY CMa. 
Most RSG are irregular variables with small 
amplitudes (type SRc or Lc), contrary to AGB stars (mostly of Mira or SRa 
type). Thus, models of mass loss for AGB stars, based on 
pulsations and radiation pressure on dust grains, are very unlikely to be applicable to RSG. 
Indeed, theoretical models have failed up to now to reproduce the observed 
mass-loss rates (Salasnish et al. 1999). Processes linked to chromospheric 
activity, convection or rotation may play an important role 
(Mallick 1993, Langer \& Heger 1998). 

In this paper we investigate the properties of the stars themselves, 
through high-resolution optical spectroscopy. The sample and 
observations are described in Sect. 2. We then present the 
results (Sect. 3) and analyse them in terms of the determination 
of the atmospheric dynamics (Sect. 4). Conclusions are given in 
Sect. 5. 

\section{Sample and observations}
The sample studied here consists of 21 late-type stars, mostly 
M type supergiants. Most of them (16 objects) are part of the 
sample studied in paper I, which itself consisted mainly of M supergiants 
found in Humphreys (1978) catalogue of the brightest Galactic stars. 
Fourteen of them have known distance and interstellar extinction, 
thanks to their OB association memberships (Levesque et al. 2005). 
The sample is presented in Table \ref{sample}. V magnitudes, extinction and 
distance moduli are taken from Lesveque et al. (2005).The latter two  
were determined on the basis of memberships to OB associations. 
The lack of data means that either the star was not identified 
as a member of such a cluster, or is not in Humphreys catalog. 

\begin{table}
\caption{Supergiants of our program. }
\begin{center}
\scriptsize
\begin{tabular}{llcccc}
\hline\noalign{\smallskip}
name & sp. type & var. type & V &  distance & A$_V$ \\
     &          &           &   &   modulus & (mag) \\
\noalign{\smallskip}
\hline\noalign{\smallskip}
HS Cas   	    & M4 Iab         &          & 9.82 & 12.0 & 2.66 \\
V466 Cas     & M2 Ib           & SRc & 8.65 & 11.0 & 1.55 \\
XX Per 	    & M4 Ib+B      &          & 8.26 & 11.4 & 1.12 \\
AD Per          & M3 Iab         & SRc & 7.90 & 11.4 & 1.91 \\
FZ Per     	    & M1 Iab         & Lc     & 7.96 & 11.4 & 1.94 \\
HD 37536    & M2 Iab         & Lc     & 6.10 & 10.4 & 1.48 \\
$\alpha$ Ori & M2 Iab      & SRc & 0.58 &  ~5.7 & 0.62 \\
UY Sct           & M2-4 Ia       &          & 8.29 &   .      &  .       \\
V336 Vul       & M2-3 I         & Lb    & 9.30 &   .      &  .       \\
BD+243902 & M1 Ia           & Lc    & 9.00 & 10.8 & 5.04 \\
BI Cyg           & M3 Ia-Iab    & Lc    & 9.33 & 11.0 & 5.11 \\
BC Cyg         & M3.5 Ia        &         & 9.97 & 11.0 & 5.58 \\
RW Cyg        & M3 Iab         & SRc & 8.13 & 10.6 & 4.49 \\
HD 203338  & M1ep Ib+B  &         & 5.66 &   .      &  .   \\
SW Cep        & M3.5 I           & SRb & 8.88 &   .     &  .   \\
$\mu$ Cep   & M2 Iab          & SRc & 4.02 & ~9.6 & 2.01 \\
VV Cep         & M2ep Ia+B  & SRc & 4.90 &   .     &  .   \\
RW Cep        & G8 Ia            & Lc    & 6.67 &   .     & 3.30 \\
ST Cep          & M2 Ib           & Lc    & 8.09 & ~9.6 & 2.32 \\
U Lac             & M4 Iab+B    & SRc & 8.70 &   .     &  .   \\
V386 Cep     & M3-4 I          &         & 8.80 &   .      &  .   \\
V582 Cas     & M4 I              &         & 8.00 &   .      &  .   \\
TZ Cas          & M2 Iab         & Lc    & 9.23 & 11.9 & 3.25 \\
\noalign{\smallskip}
\hline\noalign{\smallskip}
\end{tabular}
\end{center}
\label{sample}
\end{table}

Distance and extinction estimates rely on the assumption that the 
stars indeed belong to OB associations, which is not definitely 
established. In most cases, the properties deduced from 
this hypothesis are fully consistent (Levesque et al. 2005). 
However, some cases may be more questionable. For 
example, BC Cyg has a parallax measured by HIPPARCOS of 2.84 $\pm$ 0.87 mas, 
corresponding to a distance of at most 500 pc (1$\sigma$ error), while 
the distance modulus gives a distance of 1.5 kpc. Other peculiar cases 
are discussed hereafter. 

\subsection{Observations}
High-resolution optical spectroscopy was obtained at the Haute-Provence 
Observatory with the echelle spectrograph 
ELODIE (Baranne et al. 1996). The spectra cover the range  
from 391 nm to 681 nm, with a spectral resolving power of 42000 
corresponding to 7.0 km~s$^{-1}$. 

Automatic reduction, including bias subtraction, localisation of the 
orders, flat-field correction, and wavelength 
calibration, was achieved with the INTER-TACOS program, developed 
at the Geneva Observatory by D. Queloz and L. Weber (Queloz 1995). 
This instrument and the associated reduction package are primarily 
aimed at the search of extrasolar planets. This means that special care 
is given to wavelength calibration. Indeed radial velocities can be measured 
with an accuracy better than 0.1 km s$^{-1}$ (in the observing mode we 
used, i.e. without simultaneous observation of a calibration lamp, in contrast 
to exo-planet searches) and a great stability. 

A first set of data was obtained in August 1999 for the whole sample. 
Series of spectra were subsequently obtained for 13 of these objects 
from April 2003 to July 2004. We considered those
as the most relevant to studying atmospheric dynamics. In particular, 
we did not follow (suspected) spectroscopic binaries, 
or the faintest ones, 
to ensure a sufficient signal-to-noise ratio $(S/N)$. 
Eleven observational runs were devoted to this 
follow-up, for a total of 29 nights, resulting in an average of 7 spectra per object , and up to 11 
spectra for some objects. (Pointing limitations 
due to the equatorial mounting of the telescope, and bad weather conditions 
during the winter months, limited the time sampling in some cases.) The spectra 
have typical $S/N$ in the range 50-100 at 500 nm. 
Because of the lower sensitivity of the CCD in the blue, where most of the 
spectral lines used in the masks are located, and the fact that RSG spectra are 
very red, the $(S/N)$ of most of the observed spectral lines is an order of magnitude 
lower. However, even with an $S/N \approx 1$, the maximum error on e.g. the 
velocity broadening measurement through the cross-correlation technique 
is $\la 1$ km~s$^{-1}$,  i.e. below the typical 
microturbulence of RSG, for a spectral resolving power of 40,000 
(Queloz 1995).\footnote{All the spectra are available through the ELODIE archive 
at http://atlas.obs-hp.fr/elodie/ and through the SSA (simple spectra access) protocol 
of the Astronomical Virtual Observatory (AVO). }

\subsection{Notes on individual objects}
{\it RW Cep} - This object was included in our sample as it is 
classified as an M0 I star in Humphreys (1978). However its spectrum 
does not exhibit strong the TiO bands typical of M stars, so a spectral 
type of G8 or K Ia (Morgan \& Roman 1950) is clearly favoured. We nevertheless 
keep this star in our sample as it shows qualitatively the same 
behaviour as RSG, but with stronger asymmetries.  As it is not 
associated with any OB cluster, we adopt a distance of 840 pc (HIPPARCOS 
parallax 1.19 $\pm$ 0.54 mas) in the following analysis, 
keeping in mind that this value suffers strong uncertainty. 

\noindent
{\it Spectroscopic Binaries} - Four spectroscopic binaries, namely 
XX Per, HD 203338, VV Cep, and U Lac, were included in our initial sample. 
VV Cep and HD 203338 are in fact triple systems of the same kind 
and the binarity of XX Per and U Lac has been confirmed 
(Burki \& Mayor 1983). In the unique spectrum obtained for each 
of them, we note that they all display asymmetries and velocity 
gradients similar to those found for single RSGs. The effects of binarity (and 
perhaps accretion disks) is nevertheless probably important, as attested 
by e.g. the peculiar H$\alpha$ line profiles of VV Cep and U lac, 
which have strong emission components. We thus did not follow these 
objects. 

\noindent
{\it HD 37536} - Humphreys (1978) gives a spectral type of M2 Iab 
and an association to Aur OB1 cluster, leading to a distance of 1.4 kpc. 
But its HIPPARCOS parallax is 2.38 $\pm$ 0.97 mas (i.e. $\sim$ 420 pc). 
Our spectrum shows a strong Li 6707\AA ~line and Van Eck et al. (1998) 
indicate it is a Tc star. No significant velocity gradient was found 
for this object. It is thus more probably a misclassified AGB star and 
is excluded from the following analysis. 

\section{The tomography technique}
Velocity fields in stellar atmospheres, and convective motions in particular, 
can in principle be deduced from the study of line profiles. An  
illustration is convective blueshifts and ``C'' shaped bisectors of 
solar lines (e.g. Asplund et al. 2000). However, studies of  {\it individual} 
line profiles require observations of unblended lines at a 
resolving power of at least $10^5$ 
and a high $S/N$ ($\ga$ 100) (see e.g. Dravins 1982 for a 
review of the method and some results). The important veiling of 
spectra of late-type stars by molecular lines hampers these studies in the 
optical. 

An alternative is to use lower-resolution spectra such 
as those we obtained, together with the tomographic technique developed by 
Alvarez et al. (2001). The spectra are cross-correlated with numerical masks 
(series of holes placed at the selected line positions) that 
probe layers at different optical depths. The resulting 
cross-correlation functions (CCF) may be seen as average line profiles typical 
of a given depth of formation, with higher $S/N$ and resolution than 
the original data. Eight masks were used, designated C1 to C8, respectively, probing 
the innermost  (faint lines, with excitation potentials of $\chi_{exc} \sim 3$ eV) to 
the outermost layers (strong lines, $\chi_{exc} \sim 1$ eV). 

In the case of a hydrostatic, spherically-symmetric star, a given radial optical depth can be 
associated with a unique atmospheric height. The CCF will then appear symmetric. 
In the absence of peculiar velocity fields, such as the expansion of the atmosphere or 
the propagation of a shock wave, the CCF obtained from each mask will be shifted at  
the same velocity, corresponding to the radial velocity of the star. Any deviation from 
one of these two cases will thus probe atmospheric dynamics. 

\subsection{Reliability of the method}
The underlying assumption in this method is that all the lines that compose 
a given mask (up to $\sim$800 lines for the masks probing the innermost 
atmospheric layers) are indeed formed at the same optical depth in the 
atmosphere and thus probe similar physical conditions. A careful examination of the 
composition of the masks shows that 
the lines share a wide variety of properties, in terms of elements (and thus 
abundance and ionisation potentials), excitation potentials, 
and oscillator strengths. 

In order to check the reliability of the method, we thus examined the excitation 
conditions of the lines that make up each mask. 
The optical depth at the central wavelength of a given line is 
approximately given by 
\[   \tau_0 \, \approx \,  \frac{\sqrt{\pi}e^2}{m_e c}  \frac{\lambda_0^2}{c} 
\frac{gf}{\Delta \lambda_D} \frac{N_X}{U(T)} 10^{-\chi 5040 / T_{ex}} \]
or, for a given ion, taking into account that the Doppler width is 
$\Delta \lambda_D \, = \, \lambda v_D/c$, 
\[ \log \tau_0 \, \approx \, \log (\lambda_0) + \log gf  - \chi \theta_{ex}  + \mbox{cste}  \]
where $T_{ex}$ is the excitation temperature, $\theta_{ex} \, = \, 5040/T_{ex}$, 
$N_X$ the integrated column density of the element in the relevant ionisation 
state, $U(T)$ its partition function, $\chi$ the excitation potential, $gf$ 
the effective transition probability, and $\lambda_0$ the central wavelength 
of the transition (other symbols have their usual meaning). 
As the central optical depth of the lines in a given mask are necessarily 
similar by construction of the masks (see Alvarez et al. 2001), 
the factor $\chi /\log (gf \lambda_0)$ is representative 
of the excitation temperature, and lines selected in this way should 
probe a ``unique'' atmospheric layer. 

\begin{figure}
\centerline{\psfig{figure=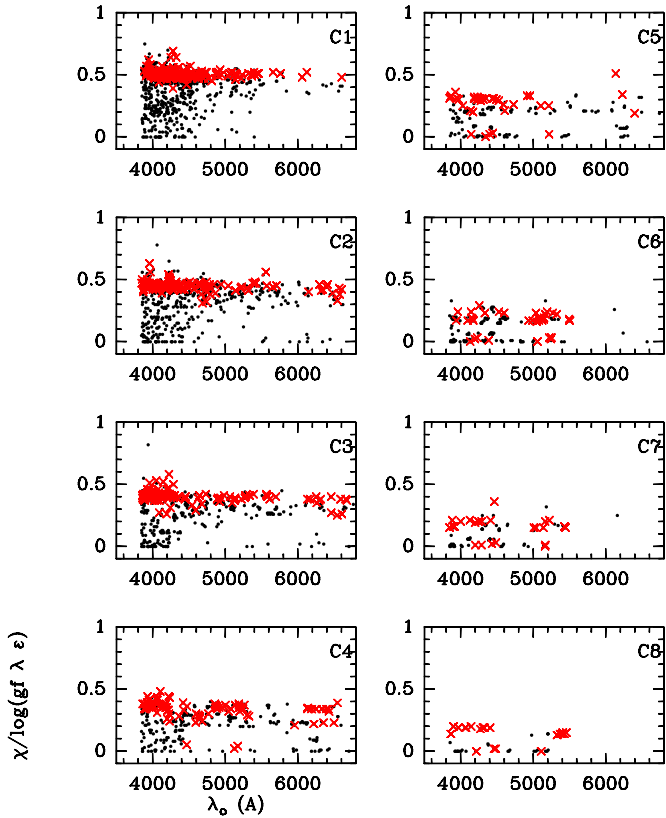,width=9cm}}
\caption[ ]{Properties of the lines within each mask defined by 
Alvarez et al. (2001) for the tomography of stellar atmospheres. 
the abundance $\epsilon$ is normalised to iron 
(i.e. $\epsilon$ = $\epsilon$(X)/$\epsilon$(Fe)). 
Fe{\sc i} lines are indicated by red crosses.  No correction for 
ionisation equilibrium are applied (i.e. the abundance of an observed 
ionisation state is supposed to  be that of the considered species). }
\label{excit}
\end{figure}

This factor, corrected for abundance effects,  is displayed in Fig. \ref{excit}, 
with special emphasis on neutral iron lines. Despite the noticeable presence  
of resonance lines in every mask, there is a clear tendency toward 
decreasing excitation temperature as one proceeds from mask C1 to mask C8. 
A strong dispersion exists however, which can be considerably reduced 
if one considers only one element in a given ionisation state (Fe{\sc i} here). 

Thus, to confirm that asymmetries found in a given CCF result from 
complex velocity fields at a given optical depth (which we are looking for) 
and are not influenced by velocity fields from different optical depths
(i.e. probed by lines with different contribution functions but grouped in the 
same mask), we computed CCF with smaller masks made uniquely 
of Fe{\sc i} lines. 

\begin{figure}
\centerline{\psfig{angle=-90,figure=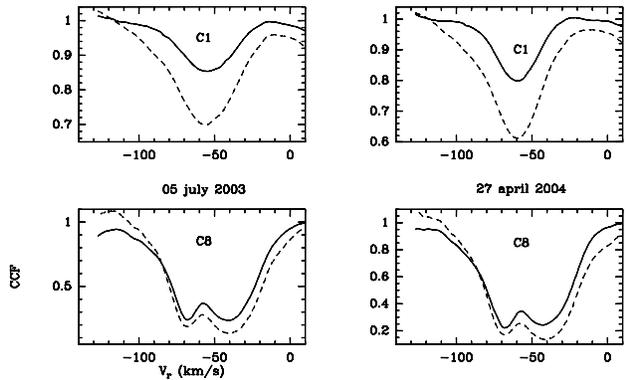,width=9cm}}
\caption[ ]{Comparison of the cross-correlation functions (CCF) obtained for 
RW Cep at two epochs (corresponding to the largest observed variations in 
velocities). The full lines are the CCF obtained with the original masks, 
the dashed lines are those obtained with Fe{\sc i} lines (see text 
for details). }
\label{testmask}
\end{figure}

Displayed in Fig. \ref{testmask} are examples of CCF obtained for one RSG at 
two epochs. We chose to show the case of RW Cep as it is the object that 
has the most asymmetric CCF, so is a good test to see if the original 
masks do not produce asymmetries of undesirable origin. One can see that CCFs 
obtained either with full masks or with Fe{\sc i} line masks 
are consistent within the increase in the 
noise level induced by the smaller number of lines in the Fe{\sc i} masks. 
In particular the central velocity of the different CCF does not change by more than  
1-2 km~s$^{-1}$, which is approximately the same as the uncertainty in the 
profile fitting (see hereafter). 
The subsequent analysis is thus based on CCFs obtained with the full masks, 
as the noise in the CCFs decreases with the increasing number of lines in the 
masks. 

We want to emphasise here that the C8 Fe{\sc i} line mask is primarily made 
of lines with excitation potentials of $\chi \sim 1-1.5$~eV. Excluding the few resonance 
lines does not affect the observed asymmetries. The CCF can thus be considered 
free of any circumstellar component, either in emission or absorption. 

\subsection{Determinations of characteristic velocities} 
In agreement with the expectation of complex atmospheric dynamics for RSG, 
we observe strongly asymmetric CCF. In order to quantify such line 
asymmetries, the traditional technique consists in determining their bisectors. 
We consider that this is not applicable here. As mentioned above, the strong 
molecular veiling observed in cool-star spectra and the spectral resolution 
of our observations renders this technique inaccurate. 
Furthermore, as many CCF exhibit two distinct minima (especially those 
corresponding to strong lines), the usual definition of the bisector becomes 
inapplicable. 

Thus, in order to characterise the atmospheric dynamics, we instead chose 
to determine characteristic velocities by fitting one or two Gaussians or 
Lorentzians to the CCFs. We measured four characteristic velocities for 
each spectrum. First, we determined an average atmospheric velocity by 
cross-correlating the spectra with a global mask, built from a synthetic 
spectrum of Arcturus (Baranne et al. 1996), which does not probe a particular 
atmospheric layer. Note that we refer here to atmospheric 
velocities, as the intrinsic stellar velocity is poorly defined, due to 
the lack of reliable indicator for these objects. 
We then adjust a single gaussian to the CCF obtained with mask C1, 
which gives access to the mean velocity of the innermost atmospheric 
layers of each star. This velocity is very similar to the global velocity, 
as almost all lines contained in masks C1 to C8 are included in the 
Arcturus mask, and lines contained in mask C1 are much more numerous 
than in any other mask and thus dominate the velocity measurement. 
We had to adjust two components to reproduce the more complex profiles 
obtained with the mask C8, and we thus measure two 
velocities for the outermost layers. Lorentzians give more satisfactory 
fits than gaussians (or a combination of both). The limitation to two 
components is questionable, as is the choice of lorentzians. 
But given the resolution, more complex adjustments would 
be meaningless. Given the number of free parameters in these 
adjustements (position, depth, and width of each component), we estimate 
that the uncertainty in the determination of these velocities is at most 2 km~s$^{-1}$ 
(i.e. the maximum difference in the velocities found with CCF obtained 
with  complete masks and with Fe{\sc i} line masks; see above). 

Naturally, even if the fits with two components are apparently satisfactory, 
this does not mean that some layers are indeed characterised by 
two distinct velocities. Some combinations of the contribution functions 
and velocity laws, since a function of optical depth (even monotonic) can lead 
to similar line profiles (e.g. Kulander \& Jefferies 1966). The 
two velocities deduced from the CCF obtained with mask C8 may 
thus be considered as indicators of the amplitude of the velocity 
dispersion in the corresponding line-forming regions. 

From these velocities we define three atmospheric velocity ``gradients'' 
\[ \delta v_{atm \, i} \, = \, v_{i}(\mbox{mask C8}) - v(\mbox{mask C1}) \;\;\;\;\; i \, = \, 1,2\]
\[ \delta v_{atm \, 3} \, = \, v_{2}(\mbox{mask C8}) - v_{1}(\mbox{mask C8}) , \]
the two velocities measured with mask C8 corresponding 
to the blue and red components. 

\section{Results}
\begin{figure}
\centerline{\psfig{figure=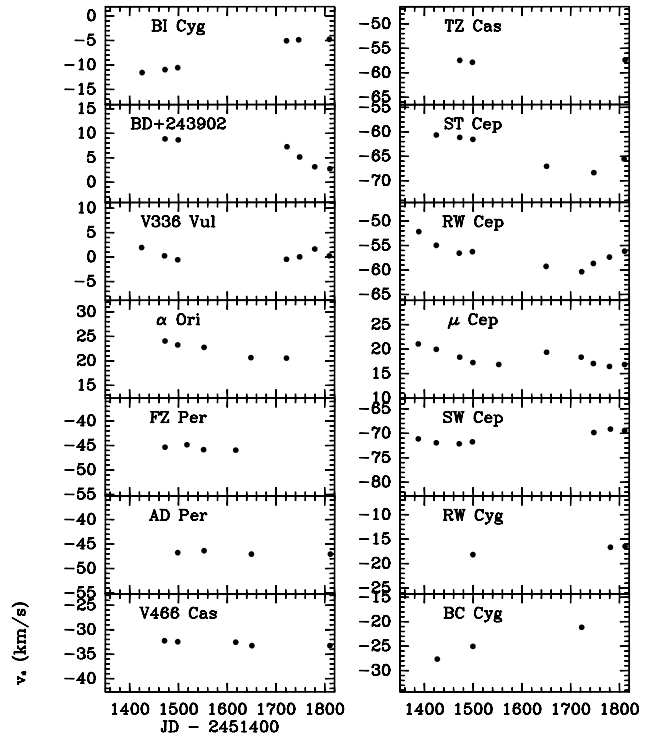,width=9cm}}
\caption[ ]{Velocity curves as a function of time. The velocities are those measured 
with the CCF obtained with the Arcturus mask. For each panel, the velocity range is fixed 
to 20 km~s$^{-1}$, in order to emphasise the relative amplitude of variations between 
each star. }
\label{stat}
\end{figure}

Figure \ref{stat} displays the observed variations of the average atmospheric velocity. 
Half of the stars have ``low'' variations 
($\delta v \, \la \, 5$~km~s$^{-1}$), while the average 
atmospheric velocity varies strongly for the others. 
Furthermore, some objects experience both local minima and maxima in 
their velocity variations during our observations, while an apparently 
continuous increase appears for two of them (BI Cyg and BC Cyg; 
but because of the scarcity of the data a minimum may have been missed). 
The amplitude of velocity variations may thus be seen as a lower limit, as 
variation timescales may exceed the duration of our monitoring. 
The best-sampled object, $\mu$ Cep, experiences variations that may be periodic, 
with a timescale of about 300 days, to be compared with the longer period of 
photometric variations ($\sim$ 860 days, Kiss et al. 2006). 

The CCFs present a wide variety of shapes (i) for each star from one 
mask to another and (ii) from one star to another. Some general 
properties can nevertheless be derived. First, the CCFs probing the 
innermost layers (mask C1) are systematically blueshifted compared 
to CCFs probing outer layers (masks C5 to C8). The asymmetry of the 
CCFs increases as one progresses from the inner to the outer layers. 
While mask C1 produces a rather symmetric profile (within the limit 
of the resolution of the observations), mask C8 produces a much 
more complex profile, with line doubling most of the time. 
In the outermost layers, when two components are found, 
one is systematically blueshifted and one redshifted relative  
to the innermost layers. Both $\Delta v_{atm1,2}$ span the same range 
of values, up to $\sim$25 km~s$^{-1}$. 

The depth, the ``central'' velocity, and the asymmetries of the CCFs are also 
time variable. Changes in the  
velocity of a given component or CCF can reach up to 13 km~s$^{-1}$ 
on a time scale of about one year. These variations are far from regular. 
Variations of 2-3 km~s$^{-1}$ within one month are sometimes observed, 
followed by an apparent stability during a few months. 

\begin{figure}
\centerline{\psfig{figure=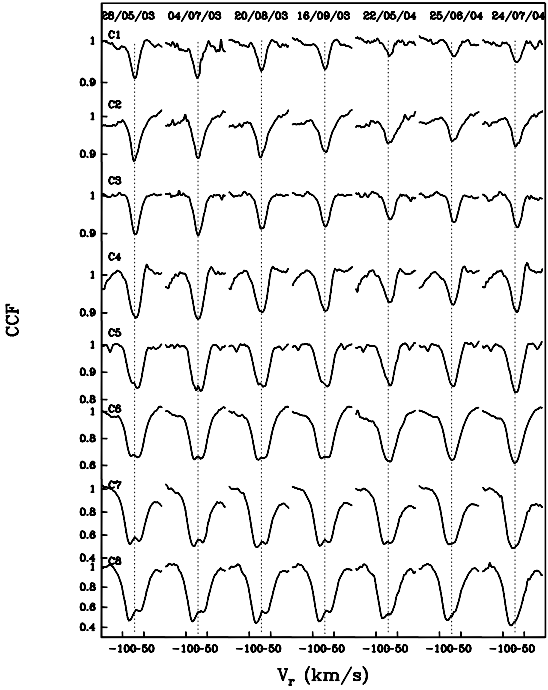,width=9cm}}
\caption[ ]{CCF profiles for SW Cep. The dashed vertical lines 
indicate the central velocity measured in mask C1 at first 
epoch (upper left profile). }
\label{sw-seq}
\end{figure}

\begin{figure}
\centerline{\psfig{figure=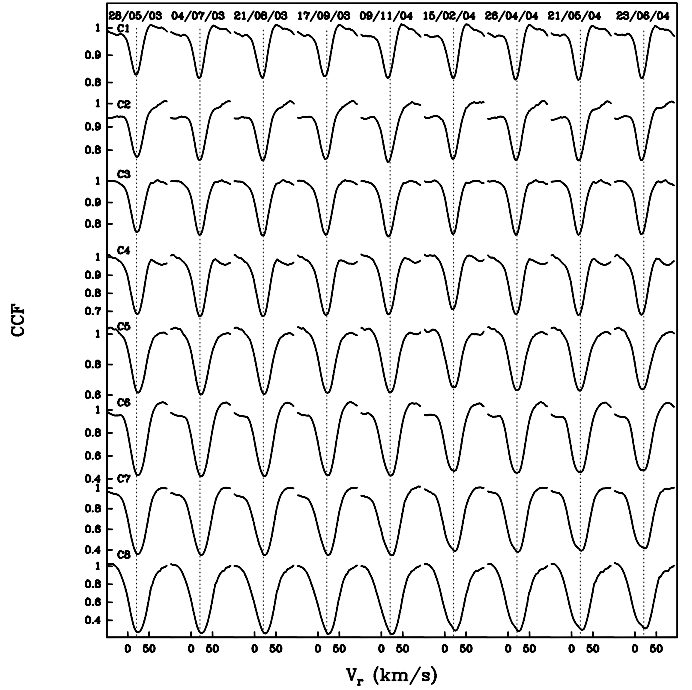,width=9cm}}
\caption[ ]{Correlation profiles for $\mu$ Cep. 
The vertical dashed lines indicate the 
central velocity measured in mask C1 at first epoch 
(upper left profile). }
\label{mu-seq}
\end{figure}

Two characteristic examples of sequences of CCFs are displayed in 
Figs. \ref{sw-seq} and \ref{mu-seq}, obtained for SW Cep and $\mu$ Cep, 
respectively. 
 In the first case, the line doubling appears rather deep in the 
atmosphere, giving rise to very broad and doubled line 
profiles in the outer atmosphere. In the second case, 
variations still exist, but seem much smoother, apparently because 
of a larger broadening of the lines, or their components. 
Even in the outer atmosphere, profiles 
appear more asymmetric rather than split into two 
components, these components being separated by less than 
$\sim$20~km~s$^{-1}$, the typical FWHM of these lines. 

\begin{figure}
\centerline{\psfig{figure=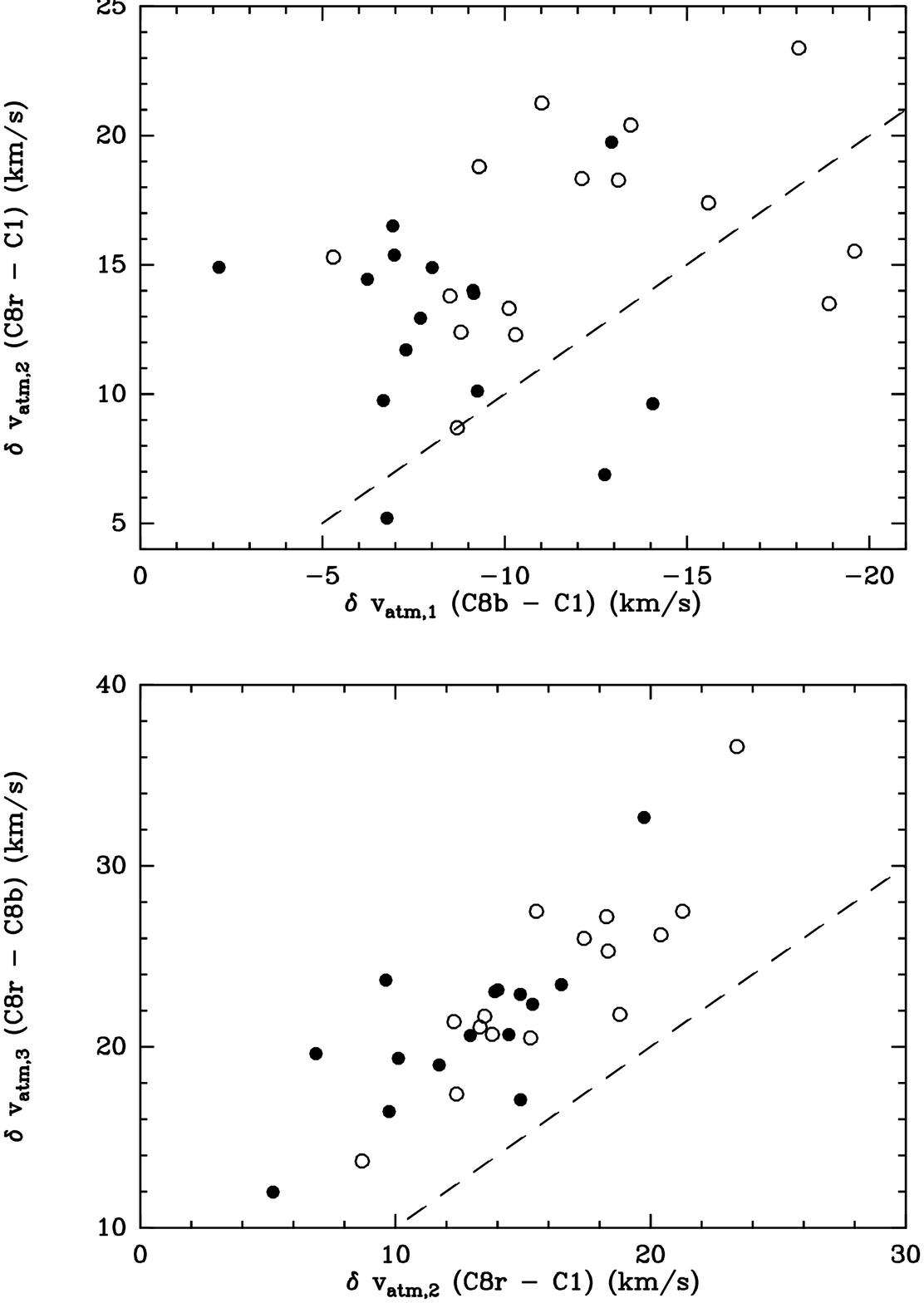,width=9cm}}
\caption[ ]{Relations between the three velocity gradients. Filled 
symbols correspond to time-averaged velocities, and empty ones to 
the maximum absolute gradients reached during our observations. 
C8b and C8r refer to the blue and red 
components in the CCF found with the mask C8. }
\label{vatm12}
\end{figure}

The relations between the different velocity gradients are displayed 
in Fig. \ref{vatm12}. A correlation between $\delta v_{atm\ , 2}$ 
(C8r-C1; which may be considered as indicative of downward motions) 
and $\delta v_{atm\ , 3}$  (C8r-C8b; convective velocity  amplitude) 
appears, while there seems to be no clear relation 
between $\delta v_{atm\ , 1}$ (upward motions) and $\delta v_{atm\ , 2}$. 
Upward and downward motions in the upper atmosphere are not symmetric 
relative to the velocity of the deeper layers. Indeed, the upward velocity is 
$v$(C8b) $\approx \; v$(C1) + 8 km~s$^{-1}$, while the downward velocity 
$v$(C8r) varies between 5 and 25 km~s$^{-1}$ w.r.t. $v$(C1). 

\begin{figure}
\centerline{\psfig{angle=-90,figure=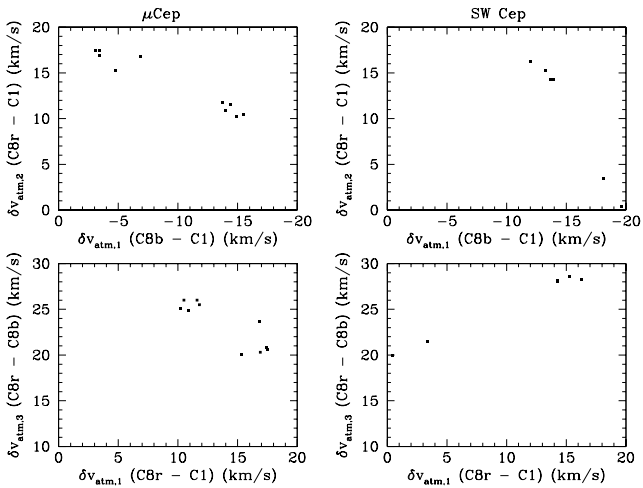,width=9cm}}
\caption[ ]{Relations between the three velocity gradients for $\mu$ Cep and SW Cep. 
Each point corresponds here to {\it one} observation. }
\label{dv-indiv}
\end{figure}

Regarding the correlation between $\delta v_{atm\ , 2}$ and $\delta v_{atm\ , 3}$ , 
the apparent lack of correlation between $\delta v_{atm\ , 1}$ and $\delta v_{atm\ , 2}$ 
may seem surprising. 
In fact, as illustrated in Fig. \ref{dv-indiv} for two objects, neither the average nor 
the extrema correspond to a situation met at any time in the star. For example, 
when $\delta v_{atm\ , 2}$ is maximum, $|\delta v_{atm\ , 1}|$ is minimum, and 
reciprocally. It is remarkable that qualitative agreement is found with simulations 
of convection in the Sun (Stein \& Nordlund 1998): downflows have higher velocities 
than upflows. 

\begin{figure}
\centerline{\psfig{angle=-90,figure=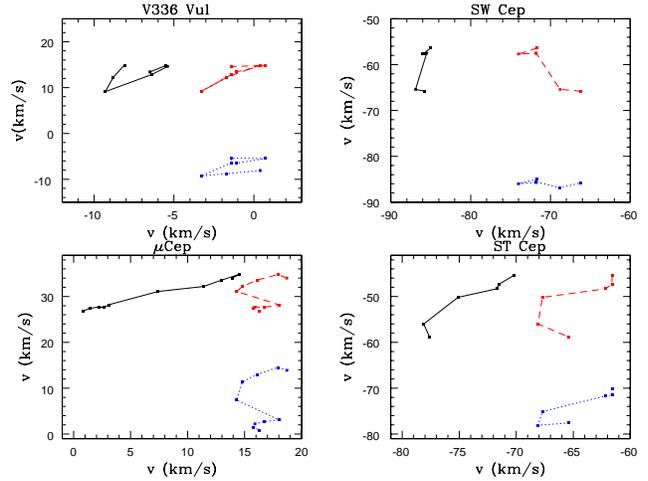,width=9cm}}
\caption[ ]{Velocity-velocity diagrams for 4 RSG. In each panel, the black, solid curve 
corresponds to v(C8r) vs. v(C8b) (ordinate vs. abscissa), the red, dashed curve to 
v(C8r) vs. v(C1) and the blue, dotted curve to V(C8b) vs. v(C1). }
\label{lissajous}
\end{figure}

The CCF profiles reveal irregular atmospheric dynamics, 
with velocities reaching supersonic values 
(the sound speed is typically $c_s \, \la \, 5$~km~s$^{-1}$ in RSG). 
Because these velocities are supersonic, one expects the propagation of shock 
waves in the atmosphere. We thus examined the possible occurrence of phase 
shift between the variations of each velocity, such as those observed in pulsating 
cool stars (e.g. Mathias et al. 1997). The velocity-velocity diagrams are shown in 
Fig. \ref{lissajous} for our objects with a good time coverage. 
No cyclic behaviour is observed, probably meaning that the atmosphere of RSG 
is not vertically stratified in a simple fashion, and shock waves are not 
spherically symmetric. 

\begin{figure}
\centerline{\psfig{angle=-90,figure=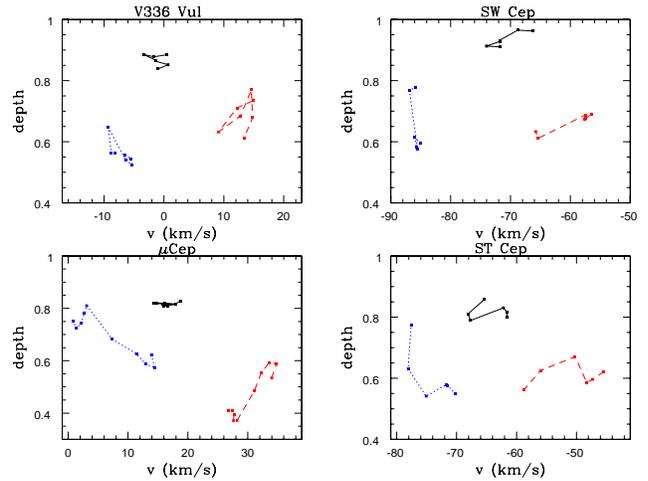,width=9cm}}
\caption[ ]{Depth as a function of velocity for the CCF obtained with mask C1 
(central black,solid curve in each panel) and the two components of the CCF obtained 
with mask C8 (left, blue, dotted and right, red, dashed curves). }
\label{depth}
\end{figure}

Examples of time variations in both depth and velocity for each component is 
shown in Fig. \ref{depth}. Here again, no cyclic or regular behaviour is found, 
confirming the irregular nature of the variations, probably associated with (or due to) 
the asymmetric atmospheric structure. 

\section{Discussion}
\subsection{Stellar parameters}
To establish the origin of the atmospheric dynamics, 
we first have to derive the fundamental parameters of our stars. 
With this aim, we used the photometry and the bolometric fluxes given in paper I 
and the distances and extinction ($A_V$) from Levesque et al. (2005)\footnote{For 
those objects not included in Levesque et al. sample, the average interstellar 
extinction of the associated cluster was used. }. Following Schlegel et al. (1998), 
we adopt $A_K \, = \, 0.112 A_V$. For $\alpha$ Ori, we adopt a distance 
of 140 pc from its Hipparcos parallax. 

The temperatures are based on the (V-K) colour and Bessell et al. (1998) 
polynomial fit to T$_{eff}$ for giants (their table 7). 
We chose the (V-K) colour as it is more sensitive to temperature than 
any visible colour (e.g. B-V) and weakly sensitive to gravity. 
Almost identical results (within 20 K) were obtained using the fit provided by 
Levesque et al. (2005), based on more recent MARCS supergiant model 
atmospheres. 
The temperature based on (V-K) for RW Cep 
is incompatible with its early spectral type. This ambiguity is probably 
at least partially due to erroneous extinction correction. (Humphreys 1978 
derived the adopted value assuming an M0 spectral type.) We thus adopt a value 
of 4200 K based on its spectral type (G8 Ia). 

\begin{figure}
\centerline{\psfig{figure=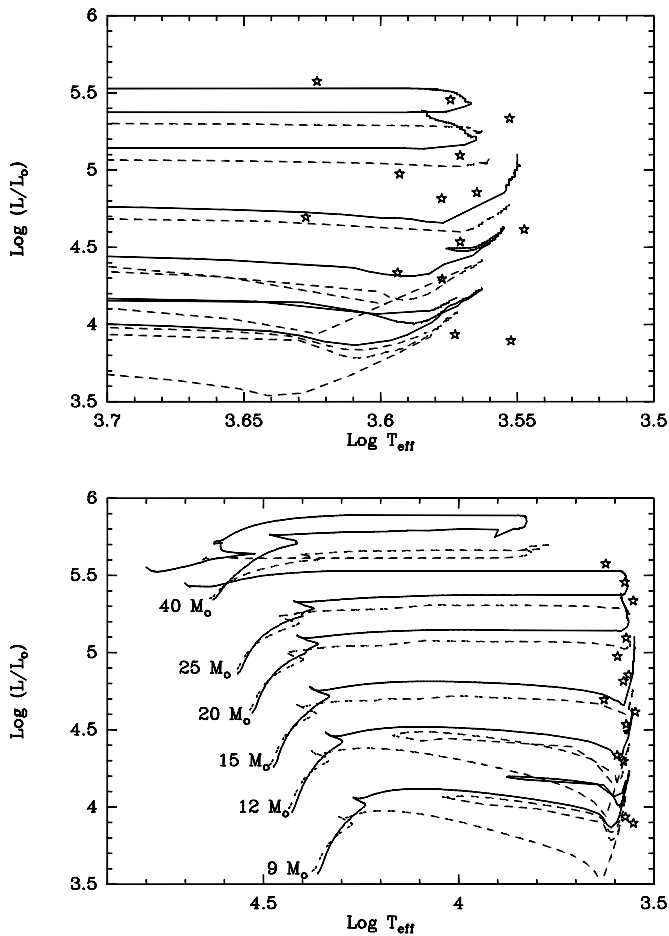,width=9cm}}
\caption[ ]{HR diagram for the sample of RSG, together with evolutionary 
tracks from Meynet \& Maeder (2003). Full-line tracks are non-rotating 
models and dashed-line tracks correspond to models with an initial 
rotational velocity of 300 km~s$^{-1}$. }
\label{hr}
\end{figure}

These temperatures allow the stars to be placed
in an HR diagram. In Fig. \ref{hr}, 
stars are plotted together with evolutionary tracks from Meynet \& Maeder 
(2003). As already noted by Levesque et al. (2005), the general agreement 
with the new evolutionary models is very good with this warmer effective 
temperature scale. We determined the initial stellar masses 
of the objects from these evolutionary tracks. They range between $\sim$10 
and 25 M$_\odot$. We keep in mind that given the inversed-square 
dependence of luminosity on distance, 
these estimates may be uncertain by at least 25\% (or 0.1 dex), which roughly 
corresponds to an uncertainty of 0.4 dex on the luminosity (or 0.2 dex on the distance, 
which is conservative, as long as the membership in OB associations is reliable). 

Following Bessell et al. (1998), we adopt M$_{bol \odot} \, = \, 4.74$ 
and T$_{eff \odot} \, = \, 5780$~K. Stellar radii are thus calculated according to 
\[ \log \left( \frac{R}{R_\odot} \right) \, = \, 8.472 - \frac{M_{bol}}{5} 
- 2 \log T_{eff} . \]
The surface gravity is then 
\[ \log g \, = \, \log (M/M_\odot) -2 \log(R/R_\odot) +  4.44 \;\; \mbox{(c.g.s.)} . \]

Using T$_{eff}$ as a typical temperature in the atmosphere, 
we determine the atmospheric pressure scale height as  
\[ H_p \, = \, \frac{P}{g\rho} \, \propto \, g^{-1}T \;\;\;\;\;\; 
H_p \, \propto \, \frac{R^2T}{M} , \] 
where $H_p$ may also be regarded as a measure of convective cell sizes 
(Schwarzschild 1975). 

Finally, as it will appear hereafter, it is useful to evaluate the density in 
the atmospheres of RSG. As no simple relation based on fundamental 
parameters exists, we examined the behaviour of the density in the photosphere 
($\tau_{Ross} \, = \, 1$) of MARCS models  (Gustafsson et al. 1975, Plez et al. 1992, 
Gustafsson et al. 2003). In the range  $3500 \la T_{eff} \la 4000$ 
and $-1 \la \log g \la 1$, we found the following relation: 
\[ \rho (\tau_{Ross}=1) \, = \, 2.6 \, 10^{-9} g^{0.6} \left( \frac{T_{eff}}{3600} \right)^{-3.5}  
\mbox{g cm}^{-3} . \]
The stellar parameters and the characteristic velocities are summarised in Table \ref{param}.

\begin{table*}
\caption{Stellar parameters and velocity gradients of our stars. }
\begin{center}
\scriptsize
\begin{tabular}{lccccccccccc}
\hline\hline\noalign{\smallskip}
name &  mass & T$_{eff}$ & log(R/R$_\odot$) & log$g$ & $\rho$ & M$_{bol}$ & $\dot{M}_{dust}$ & $<\delta v_{atm,1}>$ & $<\delta v_{atm,2}>$ & $<\delta v_{atm,3}>$ & $N^a$ \\
           & (M$_\odot$) & (K) &  & (c.g.s.) & (10$^{-9}$ g cm$^{-3}$) & & (10$^{-9}$ M$_\odot$ yr$^{-1}$) & \multicolumn{3}{c}{(km s$^{-1}$)} & \\
\noalign{\smallskip}
\hline\noalign{\smallskip}
$\alpha$ Ori  &  15  & 3780  & 2.77  &  0.08 &  2.4  &  -7.26  &   2.0   &   -9.26   &  10.13  &  19.39 &  7 \\
V466 Cas      &  12  & 3780  & 2.52  &  0.48 &  4.3  &  -6.01  &   0.2   &   -6.78   &   5.22  &  12.00 &  6 \\
AD Per	      &  12  & 3720  & 2.66  &  0.21 &  3.1  &  -6.63  &   0.8   &   -6.24   &  14.46  &  20.70 &  5 \\
FZ Per	      &  12  & 3920  & 2.51  &  0.50 &  3.9  &  -6.12  &   0.7   &   -6.68   &   9.77  &  16.45 &  6 \\
BD+243902     &  15  & 4240  & 2.63  &  0.36 &  2.4  &  -7.05  &   2.9   &   -8.02   &  14.91  &  22.93 &  7 \\
BI Cyg	      &  20  & 3720  & 2.93  & -0.12 &  2.0  &  -8.00  &   4.1   &   -6.94   &  16.52  &  23.46 &  7 \\
BC Cyg	      &  20  & 3570  & 3.09  & -0.44 &  1.5  &  -8.62  &   3.2   &   -6.98   &  15.39  &  22.38 &  4 \\
RW Cyg	      &  20  & 3920  & 2.83  &  0.07 &  2.1  &  -7.74  &   3.3   &  -12.75   &   6.90  &  19.65 &  4 \\
SW Cep	      &   9  & 3570  & 2.37  &  0.65 &  6.6  &  -5.02  &   4.6   &  -14.07   &   9.64  &  23.71 &  7 \\
$\mu$ Cep     &  25  & 3750  & 3.10  & -0.36 &  1.4  &  -8.88  &   1.5   &   -9.14   &  14.03  &  23.17 & 11 \\
ST Cep	      &   9  & 4200  & 2.24  &  0.92 &  5.4  &  -5.06  &   2.5   &   -9.16   &  13.91  &  23.07 &  7 \\
TZ Cas        &  15  & 3670  & 2.81  & -0.01 &  2.4  &  -7.35  &   3.8   &   -2.17   &  14.92  &  17.10 &  4 \\
\noalign{\smallskip}
\hline\noalign{\smallskip}
\end{tabular}
\begin{list}{}{}
\item[$^{\mathrm{a}}$] Number of observations. 
\end{list}
\end{center}
\label{param}
\end{table*}

\subsection{Atmospheric dynamics}
Velocity gradients measured on the CCF suggest both ascending and 
descending gas in the outermost layers of RSG atmospheres. 
These kinds of movements may have a convective origin. If this is 
the case, the comparison with the well-studied case of the Sun 
is instructive. As mentioned above, the pattern is in qualitative 
agreement, but the convective velocities are much higher than in the Sun. 
Furthermore, in the case of a small-scale granulation, the line profiles 
resulting from the sum of many components would produce undetectable 
line asymmetries at the resolution of our observations. 
In addition to higher velocities, we thus naturally suspect much larger 
convective cells. 

Developing an idea of Stothers \& Leung (1971), Schwarzschild (1975) 
suggested the occurrence of giant convective cells in the atmosphere 
of red (super)giants. He considered that the pressure scale height 
determines the characteristic scale of convection (in agreement 
with the mixing length theory of convection, hereafter MLT) and adopted a 
constant ratio (diameter of a granule)/(depth of a granule) $\simeq$ 3. 
Then, extrapolating solar values, he found that the entire surface 
of a red supergiant should be occupied by at most a dozen cells. Irregular 
light variations could then be attributed to changes of 
these few convective elements with time. The typical timescale would be 
$\tau \, \sim \, h/c_s$ ($h$ = typical dimension of a convective 
element, and $c_s$ = average convective velocity). With $c_s \, \sim$ 
sound velocity ($\sim$~5~km~s$^{-1}$), Schwarzschild  found 
$\tau \, \sim \, 150$~days for a typical RSG (M = 15 M$_\odot$, T$_{eff}$ = 3700 K).  
This timescale is compatible with our observed velocity 
variations (Fig. \ref{stat}). 

It is nevertheless clear that the MLT is inapproriate 
in such a scenario. A validation of this hypothesis thus requires 
detailed radiative hydrodynamics simulations. This is currently 
performed by Freytag et al. (2002). Their models confirm the occurrence 
of large convective cells, with a convective turn-over time $\sim$ 1 yr, 
about the same as Schwarzschild's timescale. 

Such supersonic convective motions could produce hot spots and 
``chromospheric'' heating (the word chromosphere being probably inappropriate, 
as this region must be very inhomogeneous and thus not spherical), 
which may be seen as blue and/or near-ultraviolet excess in the 
spectral energy distribution. This could be linked with such an excess 
observed toward $\alpha$ Ori (Carpenter et al. 1994, Harper et al. 2001) 
and maybe with that detected by Massey et al. (2005) in  a number of RSGs. 

\begin{figure}[hbt]
\centerline{\psfig{angle=-90,figure=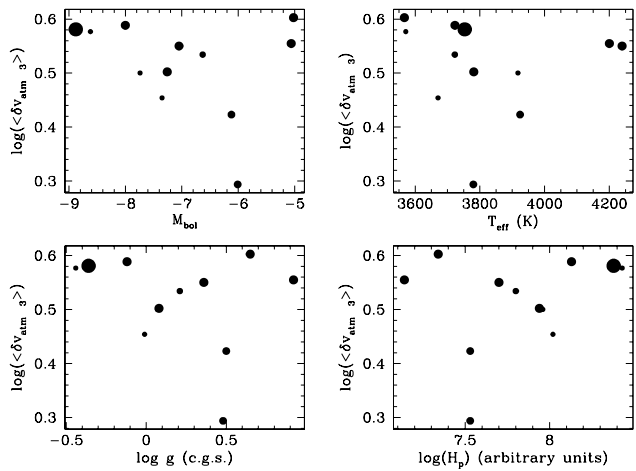,width=9cm}}
\caption[ ]{The time-averaged velocity gradient $<\delta v_{atm \, 3}>$  
as a function of bolometric magnitude, temperature, gravity, 
and pressure scale height. 
The size of the points is proportional to the number of observations, and thus indicates 
how reliable the average velocity is expected to be.   }
\label{dynamic}
\end{figure}

We now examine the relation between the atmospheric velocities and the stellar 
parameters. As shown in Fig.\ref{dynamic}, the velocity gradients themselves do 
not exhibit any convincing correlation with the stellar parameters considered here. 
This may not be surprising, as the velocity gradients do not vary by more than 
a factor $\sim 2$ within our sample, making any search for a correlation difficult. 
Rather than the atmospheric velocities themselves, it may be more meaningful 
to consider the ratio between these velocities and characteristic ones, such 
as the sound speed or the escape velocity. For example,  based on simulations 
of dwarfs and giants ($\log g \ga 2$), Freytag (2001) found an empirical 
relation between the ratio of convective-to-escape velocities and gravity: 
$v_{conv}/v_{esc} \, \propto \, g^{-1/4}$. 
But this relation actually reflects the relation between escape velocity and 
gravity ($v_{esc} = \sqrt{2gR} \propto (gM)^{1/4}$). Thus again, any search 
for a meaningful correlation is hampered by the narrow range 
spanned by velocity gradients.  

\subsection{Large convective cells or small-scale granulation ?} 
Since  the first ``images'' of Betelgeuse surface (Gilliland \& Dupree 1996) 
and the results of global simulations of RSG radiative hydrodynamics 
(Freytag et al. 2002), which both favour the occurrence of giant cells 
and/or hot spots on RSG surfaces, the reality of these features 
has been debated. In particular, Gray (2001) performed high-resolution 
($R \, \sim \, 10^5$) spectroscopic observations of Betelgeuse (91 spectra over 17 
months). He found that both line broadening and shapes are remarkably 
stable, so argues against giant convective cells. 
The changes in the strength of the lines (deeper lines, brighter star) 
could be explained by variations in the continuous opacity. 

\begin{figure}
\centerline{\psfig{angle=-90,figure=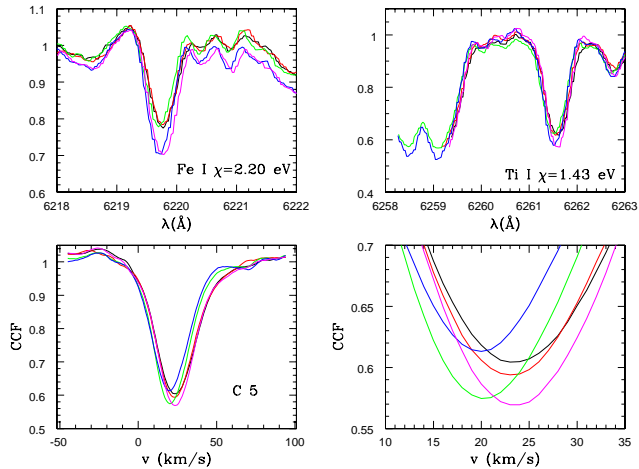,width=9cm}}
\caption[ ]{Spectroscopic observations of Betelgeuse made in 2003-2004. 
Upper panels: Two of the lines used by Gray (2001) to study the variability 
of line profiles. Lower left panel: CCF profiles calculated with the mask C5, 
which includes the Ti{\sc i} line, and Fe{\sc i} lines sharing similar properties 
with the 6219 \AA \, line (which is not included in any mask). Lower right panel: 
zoom on the central part of the CCF. } 
\label{gray}
\end{figure}

In order to see if our observations are compatible or not with these 
findings, we examined the profile variations of two of the three lines studied by 
Gray (2001) in $\alpha$ Ori spectra that we obtained (the third line, the 
$\lambda$ 6251 \AA ~line of V{\sc i}, falls between two orders of the ELODIE 
spectrograph and was thus not observed). They are shown in Fig. \ref{gray}. 
We also show the CCF obtained with the mask C5. which includes the Ti{\sc i} line. 
The Fe{\sc i} line is not present in any mask but is similar to 
other lines present in masks C4/C5. 

In contrast to Gray's assessment, we do find time-variable Doppler shifts 
and line depressions (it is not clear how radial velocity information could 
be extracted from Gray's observations). 
Asymmetries are very weak for these lines, 
but, as shown above, the most asymmetric lines have lower excitation 
potentials (see Figs. \ref{sw-seq} and \ref{mu-seq}). 
It is true however that Betelgeuse is among the least variable stars 
in our sample. 
Furthermore one cannot exclude that, during Gray's observations, 
Betelgeuse was in a rather quiet state, if thinking of quite irregular  
variations found in other RSGs. Finally we recall here that Freytag et al. 
(2002) models are not aimed at reproducing the case of 
Betelgeuse itself, but rather a typical RSG, in terms of temperature and gravity. 
The existing simulations have effective temperatures at the very lower end 
of the effective temperature scale of RSG and are probably more typical 
of very cool RSG, or AGB stars. We are now conducting  
simulations that are more appropriate for warmer RSG with B. Freytag. 

To conclude, Gray (2001) claims that ``time-variable structure in line 
profiles is a natural and necessary consequence of giant convective cells''. 
This is exactly what we observe among RSGs in general and, to a lesser 
degree of amplitude, in Betelgeuse. 

\subsection{Relation with mass loss ?}
As mentioned  in the introduction, the mass-loss process 
operating in RSG remains unknown. 
The extrapolation of the theory of the mass loss of AGB stars 
seems irrelevant, as (1) RSG have irregular, small-amplitude variations 
(and are thus not pulsating in a similar manner); (2) significant amounts of dust 
are only found at larger radii (typically $\sim$ 20 stellar radii, Danchi et al. 1994) 
so radiation pressure on dust cannot occur in the wind acceleration zone. 

Besides theory suggesting that acoustic waves (and thus 
convective motions) could initiate mass loss in RSG (Pijpers \& Hearn 1989), 
there are some observational indications that convection could play a 
key role in the mass-loss process. In particular, VY CMa, one of the RSG with 
the highest observed mass-loss rate, exhibits an asymmetric nebula, 
with arcs probably resulting from multiple ejections. Smith et al. (2001) argue 
that these arcs were most likely produced by ejection events localised on the 
stellar surface, suggesting a link with giant convective cells. 
Strong circumstellar inhomogeneities such as plumes or clumps have also been 
observed in $\alpha$ Ori shell (Plez \& Lambert 2002). 

We thus examined whether the observed velocity fields present any relation with 
mass-loss rates. Unfortunately, because of the weak emission of circumstellar 
molecular gas (paper I) and the difficulty of observing atomic gas, only 
{\it dust} mass-loss rates are available for our sample. The circumstellar 
dust-to-gas ratio is not known and may vary within RSG. 

\begin{figure}
\centerline{\psfig{figure=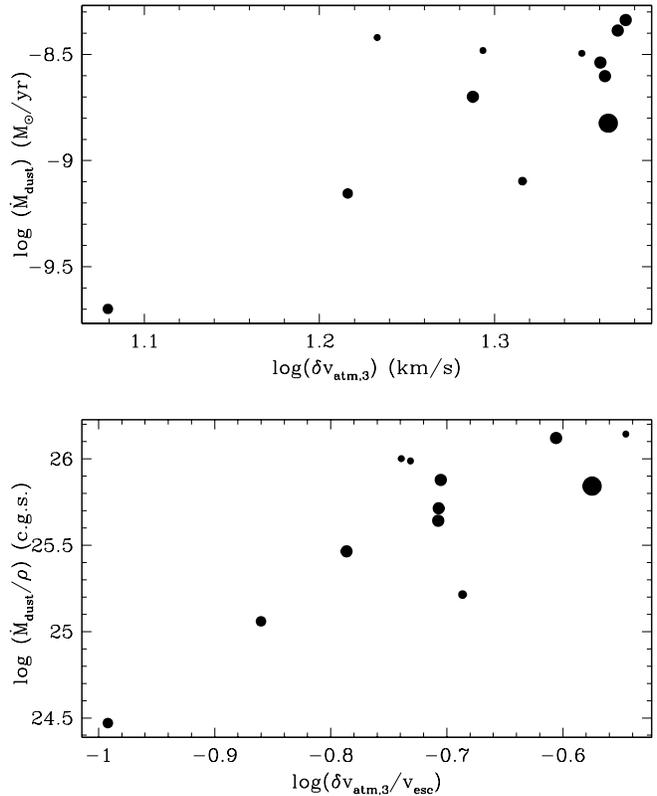,width=9cm}}
\caption[ ]{ Upper panel: Dust mass-loss rate as a function of $\delta v_{atm \, 3}$. 
Lower panel: dust rate of flow as a function of the ratio $\delta v_{atm \, 3}/v_{esc}$. }
\label{massloss}
\end{figure}

We show in Fig. \ref{massloss} the relation between the dust mass-loss rate and 
$\delta v_{atm \, 3}$, as well as between the ratio $\dot{M}_{dust}/\rho$, 
i.e. the rate of flow, and $\delta v_{atm \, 3}/v_{esc}$. A clear tendency toward higher 
rates with higher values of $\delta v_{atm \, 3}$ appears, again suggesting 
that convection may play a key role in the mass-loss process operating in RSG. 
The typical photospheric escape velocity for a RSG (with a mass of 15 M$_\odot$ and 
a radius of 500 R$_\odot$) is $\sim$100~km~s$^{-1}$. Because of the short  
mean free path of atoms in the atmosphere of RSG compared to the atmospheric extent, 
a simple evaporation, such as the Darwin's mechanism operating in planetary 
atmospheres, cannot explain the observed mass-loss rates alone. 
We recall here that, as only dust mass-loss rates are available for most of the RSG 
in our sample, no definite conclusion can be drawn as long as total mass-loss 
rates are not evaluated. Indeed, the dust condensation efficiency, and thus dust-to-gas ratio, 
may also be influenced by convective velocities and by related shock waves (through 
local density enhancements). 

In order to understand the impact convection has on the atmospheric 
structure, we may try to follow a very simple argument, based 
on the property of hydrostatic atmospheres. Additional details can be found
in Gustafsson \& Plez (1992), who discuss the impact of
turbulent pressure on density inversions and mass loss in hypergiant stars. 
We have shown above that the velocities we measured most probably reflect
the convective velocity amplitude. 
If we suppose these convective motions 
are turbulent, they give rise to a turbulent pressure of the order of 
$0.5 \rho v_{turb}^2$. We assumed further that $v_{turb}$ (in km/s)
 is a constant throughout the 
atmosphere (This may not be true, but still the gradient of turbulent
pressure will  be dominated by the density gradient) and that the atmosphere is 
isothermal. Including this pressure term in the hydrostatic
equilibrium equation leads then to 
\[  \frac{k T}{\mu m_H} \frac{d \ln\rho}{dr}  \, = \, -\frac{g}{1 + \frac{60 \mu v_{turb}^2 }{T}} = -g_{eff} ,\]
where $g$ is the surface gravity of the star, and $\mu$ the mean
molecular mass. For temperatures of 2000 to 3000~K, $g_{eff}$ is easily half the value 
of $g$, or lower, for values of $v_{turb}\approx 5{\rm kms^{-1}}$, and one tenth or less
of $g$ for $v_{turb}\approx 15{\rm kms^{-1}}$. We may thus expect that the effective gravity
of RSGs is much lower in their atmosphere than the value computed from their masses and radii.
Gustafsson \& Plez (1992) also computed the ratio of radiative to gravitational acceleration
in hydrostatic MARCS models (Gustafsson et al. 1975; Plez et al. 1992). It turns
out that the radiative acceleration on molecular lines in the outer atmospheric layers
is of the order of $g/100$ for static models at $T_{\rm eff}=3500~{\rm K}$. In a dynamic
atmosphere it may become higher, due to the combined effect of line 
desaturation by velocity fields and of cooling of the gas by adiabatic expansion.

\begin{figure}
\centerline{\psfig{figure=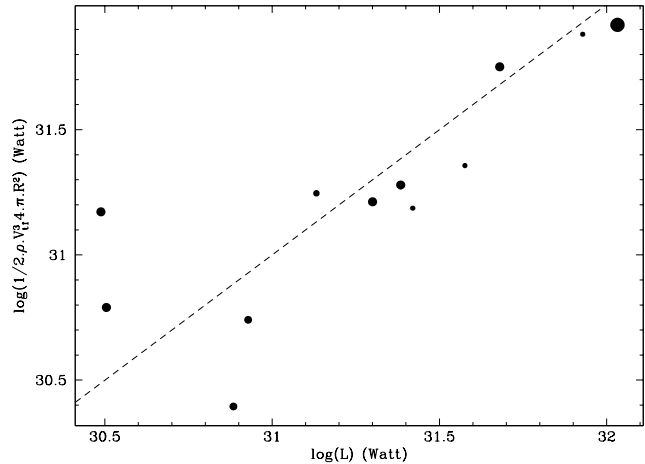,angle=-90,width=9cm}}
\caption[ ]{ Maximum rate of kinetic energy flux injected by turbulent motions as a function 
of stellar luminosity. }
\label{kinetic}
\end{figure}

It is possible to estimate the maximum rate of kinetic energy flux injected by turbulent 
motions as 
\[ \frac{dE_c}{dt} \, = \, \frac{1}{2} \rho v_{turb}^3 \, 4\pi R_\star^2 . \] 
For the stars in our sample, using our determination of $<\delta v_{atm,3}>$ for $v_{turb}$, 
we find typical numbers between $10^{30.5}$ and $10^{32}$ W. We find that the power 
of convection as measured by our indicators is about equal to the luminosity of the stars 
(Fig. \ref{kinetic}). The wind power 
is $\dot{M}(v_{esc}^2 + v_\infty^2) \, \approx \, \dot{M}v_{esc}^2$,  as $v_\infty$ is about 
3 times lower than $v_{esc}$. For a gas-to-dust ratio of 400, this leads to 
$10^{26}-10^{27.5}$ W. There is thus plenty of power available from convective motions 
to initiate the wind. 

We therefore suggest that mass loss is initiated in RSG by the combined effect
of a vigorous convection and radiative pressure on molecular lines.
Detailed calculations for radiative-hydrodynamical models of RSG 
will be presented in a forthcoming paper. 

Other and/or additional 
origins of mass loss, in particular magnetic fields, are not excluded. 
In AGB stars, large-scale fields may be strong enough to form 
magnetic cool spots, which could regulate dust formation, hence 
mass loss. Coupling Freytag's model for $\alpha$ Ori with a MHD code, 
Dorch (2004) found that surface magnetic fields up to $\sim$ 500 
Gauss can be produced, in agreement with recent measurements by 
Vlemmings et al. (2005). These strong magnetic 
fields could be generated by convective motions, through 
$\alpha \omega$-dynamo (Blackman et al. 2001). 

\section{Conclusion}
High-resolution spectroscopic time-series observations of a
sample of RSG, analysed with a tomography technique, reveal 
variable velocity fields in the atmosphere of RSG, probably 
of convective origin. These velocities are supersonic and vary 
with time scales of a few 100 days. 
The strong line asymmetries 
generated by these velocity fields indicate that convection must consist of giant 
cells, as also suggested by recent radiative hydrodynamics modelling (Freytag 
et al. 2002). We find a behaviour of upward, downward, and horizontal velocities 
similar to that characterising solar convection. Such constraints will be used to 
test new RHD simulations. 

 These convective motions could play an important role in the mass-loss process 
operating in RSGs, because of the strong decrease in effective gravity 
due to turbulent pressure. 

Further improvements in our understanding of these puzzling objects 
will be presented in a future paper (Chiavassa et al., in preparation), 
which will compare radiative hydrodynamic simulations with these observations.

\begin{acknowledgements}
We are grateful to Rodrigo Alvarez for providing his correlation 
procedures and his numerical masks, and to Bernd Freytag and Nicolas 
Mauron for fruitful discussions. We also thank the referee for useful comments. 
We acknowledge the staff of the Haute-Provence Observatory for 
help during the observations. 
\end{acknowledgements}

\end{document}